\documentclass[letterpaper, 10 pt, conference]{ieeeconf}  

\IEEEoverridecommandlockouts                              
\usepackage{graphicx} 
\usepackage{comment}

\title{Multi-Cloud Resource Management using Apache Mesos for Planned Integration with Apache Airavata }

\usepackage{authblk}

\author[1]{Pankaj Saha}
\author[1]{Madhusudhan Govindaraju}
\author[2]{Suresh Marru}
\author[2]{Marlon Pierce}
\affil[1]{Department of Computer Science, SUNY Binghamton}
\affil[2]{Science Gateway Group, Indiana University}

\begin{document}

\maketitle
\thispagestyle{empty}
\pagestyle{empty}

\begin{abstract}

We discuss initial results and our planned approach for incorporating Apache Mesos based resource management that will enable design and development of scheduling strategies for Apache Airavata jobs so that they can be launched on multiple clouds, wherein several VMs do not have Public IP addresses. We present initial work and next steps on the design of a meta-scheduler using Apache Mesos. Apache Mesos presents a unified view of resources available across several clouds and clusters. Our meta-scheduler can potentially examine and identify the cases where multiple small jobs have been submitted by the same scientists and then redirect job from the same community account or user to different clusters. Our approach uses a NAT firewall to make nodes/VMs, without a Public IP, visible to Mesos for the unified view.

\end{abstract}

\section{INTRODUCTION}

Resources allocated on the NSF clouds typically consist of a few public IP nodes, and several nodes that can only be internally routed. Campus clusters, such as the one at Binghamton University, use firewalls to prevent the starting up of server side sockets/services. The off-the-shelf packaging of Apache Mesos~\cite{HindmanMesos:Center} and Mesosphere Marathon~\cite{Marathon:DC/OS} works only when all the participating nodes have public IPs or they belong to the same network. To address this limitation, we are working on adapting the packaging and configuration of the Apache Mesos and Marathon tools to allow efficient selection of resources from different clusters, and also allow applications to span clusters and clouds. Heavyweight, commercial job schedulers are too costly for our target use case. With this project, gateways will have the opportunity to add an open source Cloud and Big Data resource manager to reduce overall wait time for bursty jobs and make the cluster and cloud resources utilized efficiently in the increasingly heterogeneous computing environment. 

\begin{figure}[!ht]
  \begin{center}
    \includegraphics[width=0.49\textwidth]{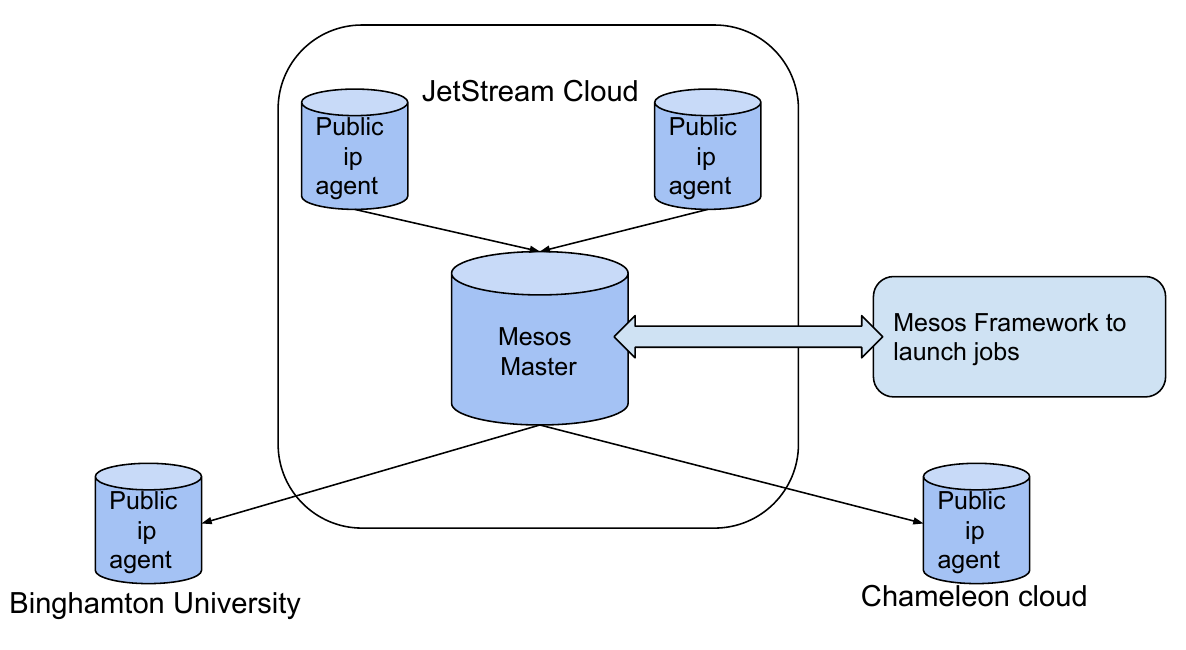}
    \caption{Shows how mesos agents from different network are connected to mesos master to create a cluster of nodes from different clouds.}%
 \label{fig:multi-cluster}%
 \end{center}
\end{figure}

\section{Target Use Cases}

Due to the high demand of many computing grids and clusters, a queued job may take several hours or even days to begin executing. This is especially cumbersome for large jobs which require a high amount of processing power because the scheduler needs to wait until enough resources are available as well as ensure a fair distribution of time. A direct consequence of high demand is a restriction on the number of jobs submitted to each cluster. As a result, users are stuck in a catch-22: the most efficient way to complete a task is to break it down into smaller jobs, but submitting many small jobs will often conflict with the cluster’s restrictions and policies, such as number of jobs that can be queued. This is especially problematic for gateways using community accounts. It is important to note that these restrictions are on a per-cluster basis, so they must be addressed for each compute resource. 

\section{Experimental Setup and Initial Results}

To setup a meta scheduling infrastructure we have combined three cloud networks from Chameleon~\cite{Mambretti2015NextSDN}, Jetstream~\cite{Stewart2016Jetstream}, and a campus cluster at Binghamton University. Mesos agents are currently running on each of the cloud nodes with a Public IP associated with them, offering their resources (CPU, memory and disk ) to the Mesos master, which can reside on any one of the cloud nodes. The Mesos master launches jobs on the agents that are chosen using the Marathon framework. By default, internal nodes, with no public IP, are not accepted as part of public Mesos cluster due to their inability to open server side sockets. To address this restriction, we have setup a NAT firewall that maps services on internal nodes to a Public IP and port. Our goal is that the end users of this project should be able to submit jobs through the Airavata PGA client \cite{Saha2016IntegratingMesos}\cite{ApacheAiravata}, which in turn will communicate with the Airavata server and get access to a set of Mesos cluster nodes that may be distributed across clouds. 
\begin{figure}[!ht]
  \begin{center}
    \includegraphics[width=0.49\textwidth]{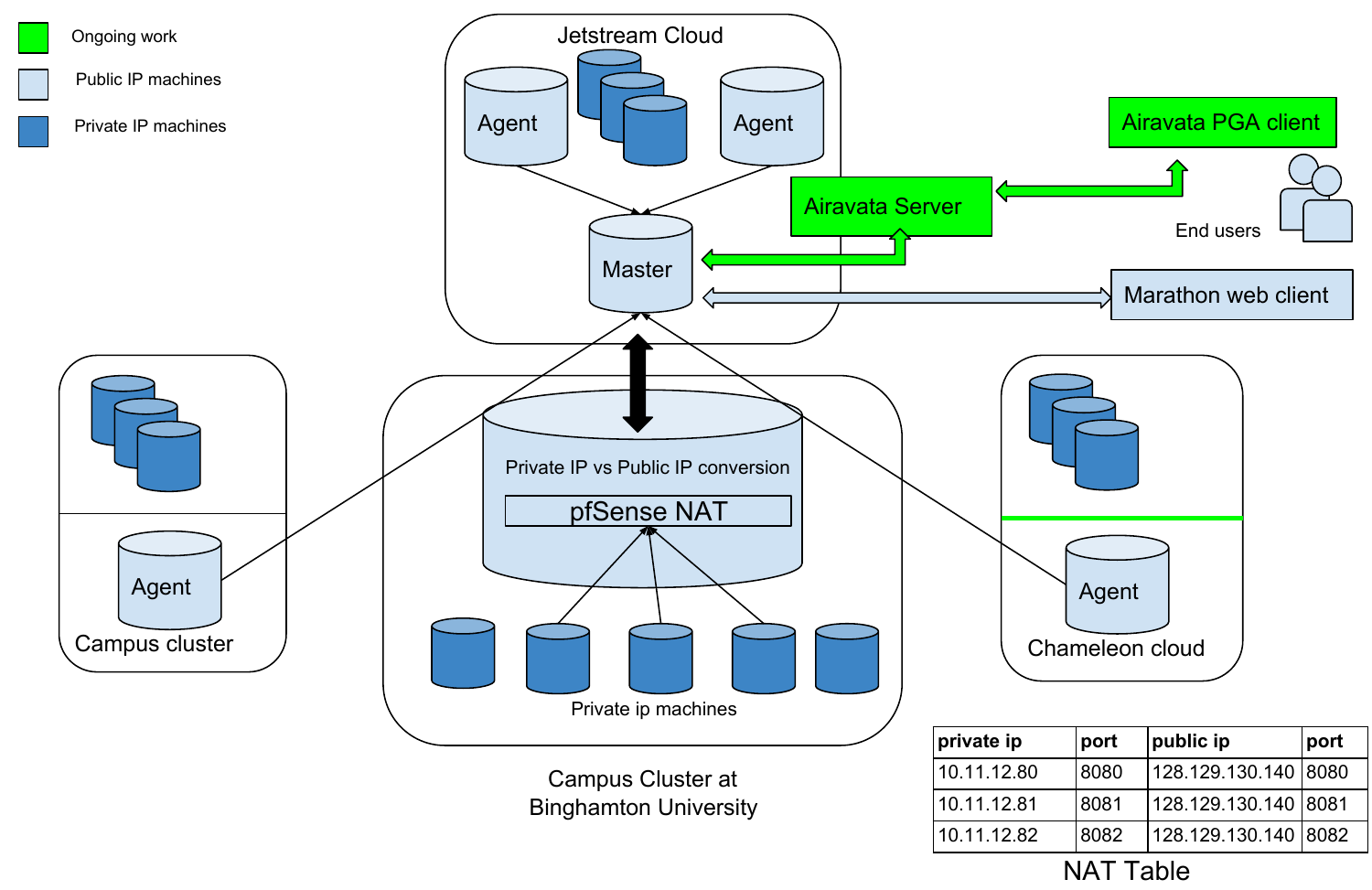}
    \caption{Apache Mesos based job launch across three clouds/clusters with a mix of Public IPs and internally accessible nodes. Jobs can span clouds and Queue Backfilling algorithms can be specialized for communities accounts.}%
 \label{fig:hybrid_cluster}%
 \end{center}
\end{figure}
Figure 1 shows the current capability of our setup wherein we can submit jobs using Mesos to different clouds and clusters. Figure 3 shows the screenshot from Marathon’s interface with the status of the jobs that have been launched. All the jobs though have been launched on nodes in the clouds and clusters that have a Public IP. Figure \ref{fig:hybrid_cluster} shows our ongoing work on using a NAT to add internal cloud nodes, without Public IPs, to the Mesos cluster. 
\begin{figure}[!ht]
  \begin{center}
    \includegraphics[width=0.49\textwidth]{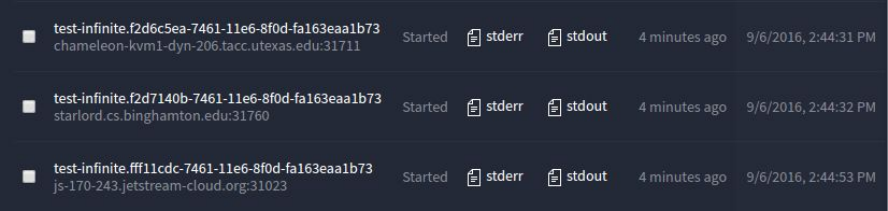}
    \caption{Screenshot showing Mesosphere Marathon launching jobs on Chameleon, Jetstream, and our Campus cluster node. While the nodes are on three different clouds/clusters, the nodes all have public IP addresses}%
 \label{fig:multiple_cloud}%
 \end{center}
\end{figure}

\begin{figure}[!ht]
  \begin{center}
    \includegraphics[width=0.49\textwidth]{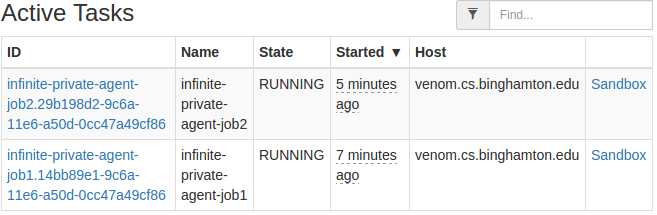}
    \caption{Screenshot showing active tasks on Mesos agents behind campus firewall}%
 \label{fig:private-agents}%
 \end{center}
\end{figure}

\begin{figure}[!ht]
  \begin{center}
    \includegraphics[width=0.49\textwidth]{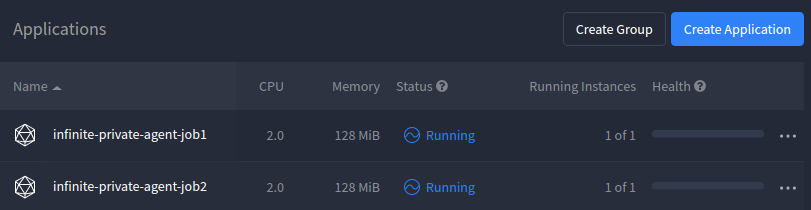}
    \caption{ Screnshot shows Mesosphere Marathon launching jobs on Mesos agents behind campus firewall}%
 \label{fig:firewall}%
 \end{center}
\end{figure}

Figure 4 and 5 shows, that we have successfully modified the Mesos installation and used NAT so that Mesos agents, without Public IPs. So, now agents with internal private IP address can also contribute their resources to a public Mesos cluster and receive job requests to execute them.

\section{Next Steps}

We plan to explore how user defined policies for fair share scheduling can be implemented, instead of the default scheduling policies for jobs, for all the job requests on the shared compute resources. The current policies on shared resources, such as XSEDE, do not currently apply fairly to problems such as parametric sweep executions. Apart from enforcing user defined fair use policies based on number of jobs, our meta-scheduler solution will seek to apply the limits by actual usage across multiple dimensions such as duration, number of cores, and memory.
Our NAT setup on a campus cluster currently makes the internal nodes visible for the Mesos Master, which in turn can take into account all the available nodes, with or without Public IPs, to determine where to launch the next job. We will explore the use of the networking APIs exposed by Chameleon and Jetstream to implement a similar NAT on these cloud infrastructures.
This will also helps us design custom backfilling algorithms that can run the jobs out of order.

\bibliography{Mendeley.bib}
\bibliographystyle{IEEEtran}

\end{document}